**Ferroelectric Domain Patterning Controlled Schottky Junction State in Monolayer MoS$_2$**

Zhiyong Xiao[1,2], Jingfeng Song[1,2], David K. Ferry[3], Stephen Ducharme[1,2], Xia Hong[1,2*]

[1] Department of Physics and Astronomy, University of Nebraska-Lincoln, NE 68588-0299, USA

[2] Nebraska Center for Materials and Nanoscience, University of Nebraska-Lincoln, NE 68588-0299, USA

[3] School of Electrical, Computer, and Energy Engineering, Arizona State University, Tempe, Arizona 85287-5706, USA

[*]xia.hong@unl.edu

**ABSTRACT**:

We exploit scanning probe controlled domain patterning in a ferroelectric top-layer to induce nonvolatile modulation of the conduction characteristic of monolayer MoS$_2$ between a transistor and a junction state. In the presence of a domain wall, MoS$_2$ exhibits rectified *I-V* that is well described by the thermionic emission model. The induced Schottky barrier height $\Phi_B^{eff}$ varies from 0.38 eV to 0.57 eV and is tunabe by a SiO$_2$ global back-gate, while the tuning range of $\Phi_B^{eff}$ depends sensitively on the conduction band tail trapping states. Our work points to a new route to achieving programmable functionalities in van der Waals materials and sheds light on the critical performance limiting factors in these hybrid systems.

Monolayer (ML) transition metal dichalcogenide (TMDC) MX$_2$ (M = Mo, W; X = S, Se, Te) possesses atomic thickness, highly tunable electronic states, and direct band gaps of 1-2 eV [1], making them a versatile playground for imposing nanoscale band structure design [2] and building novel homo- and hetero-junctions for nanoelectronic [3-6] and optoelectronic applications [7-9]. TMDC-based Schottky and *p-n* junctions have previously been achieved by exploiting multiple local gates [9, 10], electric-double-layer gating controlled bipolar carrier injection [11], contact work function engineering [12, 13], or exploring van der Waals heterostructures [14-16], where the device functionalities cannot be altered after fabrication. On the other hand, the ability to create well-defined but reconfigurable potential barriers at the nanoscale can be utilized to disentangle the intricate interplay of various impurity sources and the dielectric environment in determining the electronic and optical response of TMDC in the junction state, as well as achieving multifunctional operations on a single two-dimensional (2D) platform.

A promising approach to locally defining the potential profile and producing a programmable junction state in TMDCs is to capitalize on the nonvolatile, switchable polarization field of a ferroelectric gate. The ferroelectric field effect has previously been intensively investigated in layered van der Waals materials such as graphene and TMDCs for building novel logic and memory devices [17-20], tunnel junctions [21], sensors [22-24], and plasmonic and optoelectronic applications [25-28]. However, the unique opportunity offered by nanoscale polarization control, which can lead to local carrier density modulation in the 2D channel, has yet to be explored. In this study, employing domain patterning in an ultrathin ferroelectric polymer top layer via scanning probe microscopy, we have achieved for the first time programmable transistor and Schottky junction states in a single channel of ML MoS$_2$. In the mono-domain states, we observed ohmic source-drain current-voltage ($I_{DS}$-$V_{DS}$) relation for both the polarization up ($P_{up}$) and down ($P_{down}$) states. Modeling the transfer characteristics shows that the free electron band mobility of the sample is limited by charged impurity (CI) scattering, which is not affected by polarization switching. In the presence of a ferroelectric domain wall (DW) perpendicular to the current direction (denoted as the half $P_{up}$-half $P_{down}$ state), the channel exhibits rectified *I-V* that can be well described by the thermionic emission model, with the induced Schottky barrier height $\Phi_B^{eff}$ further tunable by a SiO$_2$ global back-gate.

We fabricated ML MoS$_2$ field effect transistor (FET) devices sandwiched between a 300 nm SiO$_2$ global back-gate and a ferroelectric top layer (17.8±0.7 nm) of poly(vinylidene fluoride-trifluoroethylene) [P(VDF-TrFE)] (Fig. 1(a)), where the copolymer film was deposited using the Langmuir-Blodgett (LB) technique [29-31]. Unlike the thick P(VDF-TrFE) films prepared by spin-coating [17, 18, 22, 26], the ultrathin LB films possess smooth surfaces (~1 nm surface roughness) and low coercive voltage (< 10 V), making it possible to write and image ferroelectric domains with nanoscale precision via the conductive probe atomic force microscopy (AFM) and piezo-response force microscopy (PFM) [37, 38]. The reported electrical results were based on 6 MoS$_2$ samples, denoted as D1-D6.

Figure 1(b) shows the room-temperature transfer characteristic of sample D2 gated through SiO$_2$ (Fig. 1(c)). With increasing back-gate voltage $V_{BG}$, $I_{DS}$ exhibits an exponential growth at low $V_{BG}$ followed by a quasi-linear gate-dependence beyond a transition voltage (denoted as $V_t$), signaling the shift of the Fermi level $E_F$ from well residing in the bandgap to close to the conduction band edge. We extracted the field effect mobility of the sample from the linear region of $I_{DS}(V_{BG})$ as $\mu_{FE} = \frac{1}{e\gamma}\frac{L}{W}(dG/dV_{BG})$, where $\gamma = 7.2\times10^{10}$ cm$^{-2}$V$^{-1}$ is the gating efficiency of 300 nm SiO$_2$, $G = I_{DS}/V_{DS}$ is the channel conductance, and $L$ ($W$) is the channel length (width). As there is no metal-electrode on P(VDF-TrFE), the presence of the ferroelectric top layer does not change $\gamma$, as demonstrated in previous studies of SiO$_2$-gate graphene with ice [39] and HfO$_2$ [40] top-layers. In the as-deposited state, the sample exhibits $\mu_{FE}$ of 6.7 cm$^2$V$^{-1}$s$^{-1}$ and $V_t$ of -20 V.

To control the out-of-plane polarization of the copolymer, we apply a constant bias voltage higher than the coercive voltage to a conductive AFM tip while it is scanning on the sample surface in the contact mode (Fig. 1(a)). We first switched the copolymer to the uniform $P_{down}$ state by writing the entire channel area with a tip bias $V_{tip}$ of +10 V while keeping MoS$_2$ grounded. The PFM phase image (Fig. 1(d)) shows that P(VDF-TrFE) on top of the device area has been uniformly polarized, while the portion on top of the SiO$_2$ substrate remains unpoled. In the $P_{down}$ state, electrons are accumulated in MoS$_2$, which shifts $E_F$ towards the conduction band edge, yielding a nonvolatile offset in the doping level. The resulting transfer curve shifts to the negative $V_{BG}$ direction, with corresponding $\mu_{FE}$ of 4.7 cm$^2$V$^{-1}$s$^{-1}$ and $V_t$ of -38 V.

We then polarized the ferroelectric top-layer to the uniform $P_{up}$ state (Fig. 1(e)) by scanning the entire sample area with $V_{tip}$ = -10 V, depleting electrons from MoS$_2$. The resulting $I_{DS}$-$V_{BG}$ curve shifts to the positive $V_{BG}$ direction with $V_t$ = -12 V. The slightly lower $\mu_{FE}$ of 3.3 cm$^2$V$^{-1}$s$^{-1}$ is due to the reduced screening of CIs in the depletion state. The scanning probe controlled polarization switching is nonvolatile and fully reversible. The third AFM writing with $V_{tip}$ = +10 V poled the copolymer back to the uniform $P_{down}$ state, and the resulting transfer curve of MoS$_2$ overlaps with that of the first poling. The switching ratio between the high ($P_{down}$) and low ($P_{up}$) conductance states reaches ~450 at $V_{BG}$ = -25 V.

In previous studies of ferroelectric-gated TMDC, despite the tremendous progress in developing various nanoelectronic/optoelectronic devices [17-22, 25, 26], in-depth understanding is yet to be gained on how the ferroelectric layer affects the mobility of the TMDC channel, e.g., regarding the presence of a charged interface, the remote interfacial polar (RIP) phonon, and the polarization reversal. To assess these ferroelectric-specific factors, we have quantitatively modeled the transfer characteristics of MoS$_2$ in both polarization states. The temperature-dependence of $I_{DS}$ shows that $E_F$ for both polarization states lies in the band gap ($E_g$) for the entire gating-range [31]. It is thus important to consider in the modeling the conduction band tail trapping states within the gap region as well as the localized states above the conduction band edge, which have been attributed to the charged impurities such as S vacancies in MoS$_2$ and CIs from the dielectric environment [6]. To account for these midgap states, we incorporated the density of states (DOS) distribution in MoS$_2$ proposed in Ref. [6]:

$$D_n(E) = \begin{cases} \alpha D_0 \exp\left[-\dfrac{E-E_D}{\varphi}\right] & E_D - \dfrac{E_g}{2} < E < E_D \\ D_0 - (1-\alpha)D_0 \exp\left[-\dfrac{E-E_D}{\varphi'}\right] & E > E_D \end{cases} \qquad (1)$$

Here $D_0 \sim 3.9 \times 10^{14}$ eV$^{-1}$cm$^{-2}$ is the 2D DOS for crystalline MoS$_2$, $E_D$ is the conduction band edge, $\varphi$ is the characteristic depth of the band tail, and $\alpha$ and $\varphi'$ are fitting parameters. Using Eq. (1), we first modeled the $V_{BG}$-dependence of the total carrier density $n_{total}$ and the free electron density $n_{free}$ for sample D3 in both polarization states at $V_{BG} > V_t$ (Fig. 2(a)) [31]. Here $n_{total}$ includes both trapped/localized electrons and $n_{free}$, with $\Delta n_{total} = \gamma \Delta V_{BG}$. For $n_{free}$, we only consider the extended states above the mobility edge energy $E_M$ with respect to the conduction band edge:

$$n_{free}(300K) = \int_{E_M}^{\infty} D_n(E) f(E, 300K) \, dE, \qquad (2)$$

where $D_n(E)$ is given by Eq. (1) and $f(E)$ is the Fermi-Dirac distribution. We used the parameters of $\varphi = 110$ meV and $E_M = 10$ meV similar to those reported in Ref. [6], as our samples possess similar mobility values.

Figure 2(b) shows the free electron band mobility $\mu_{band} = \sigma/n_{free}e$ as a function of $n_{free}$, which reveals two important points. First, even though the field effect mobility in this region is a constant (i.e., $I_{DS}(V_{BG})$ is linear), the band mobility associated with the free electrons actually increases linearly with increasing $n_{free}$ for both polarization states, reaching ~10 cm$^2$V$^{-1}$s$^{-1}$ at $n$ = 1x10$^{11}$ cm$^{-2}$. This suggests CI scattering rather than phonon scattering as the dominant mobility limiting mechanism [41], since the increased density can enhance the screening of the CIs, resulting in higher mobility. Using Boltzmann transport theory with relaxation time approximation, we find this mobility level corresponds to a CI density $N_i$ ~3x10$^{12}$ cm$^{-2}$ [31], comparable with the value obtained in MoS$_2$ sandwiched between SiO$_2$ and HfO$_2$. It further confirms that the parameters for the band tail states and mobility edge in Ref. [6] can give a reasonable description of the charge trapping/localization in our samples. Our device mobility extracted from the gap region is also consistent with the high density mobility values reported in literature for MoS$_2$ on SiO$_2$ [6, 42-44], as discussed in the Supplementary Materials [31]. At this low density range, scattering from intrinsic acoustic and non-polar optical phonons in MoS$_2$ is negligibly small. Despite the high dielectric constant of P(VDF-TrFE) ($\varepsilon_{PVDF} = 10$), its RIP phonon also does not play a critical role in limiting the mobility of MoS$_2$, in stark contrast with other high-$k$ dielectrics such as HfO$_2$ and Al$_2$O$_3$ [41]. While P(VDF-TrFE) can provide effective dielectric screening to the RIP modes of SiO$_2$, its most active optical phonon modes are at $\hbar\omega$ = 63 and 109 meV [45], and the corresponding RIP modes would not contribute as strongly as that of SiO$_2$ at room temperature [46]. Second, at the same $n_{free}$, the sample possesses similar mobility values for both polarization states, indicating that switching the polarization direction does not introduce additional scatterers. This is understandable since the switchable polarization originates from the crystalline fraction of P(VDF-TrFE), which imposes a periodic polarization field that does not scatter electrons.

We next explored the possibility of creating a potential barrier in the $MoS_2$ channel by writing a ferroelectric DW perpendicular to the current direction. In the mono-domain states, despite the large current modulation, we observe linear, ohmic-like $I_{DS}$-$V_{DS}$ characteristics for both polarization states within $V_{DS} = \pm100$ mV (Fig. 3(a)) over the entire $V_{BG}$ range for all samples investigated. In sharp contrast, when P(VDF-TrFE) was patterned into two domains with opposite out-of-plane polarization directions (Fig. 3(b) inset), at the same $V_{BG}$ of -10 V, sample D4 exhibits rectified nonlinear conduction with much weaker variation in $I_{DS}$ at reverse-bias (Fig. 3(b)). Such nonlinearity can be further tuned by $V_{BG}$. Figure 3(c) shows the color contour of $I_{DS}$ normalized to the value at $V_{DS} = +100$ mV versus $V_{DS}$ and $V_{BG}$ for this sample, which exhibits a clear evolution between two distinct behaviors with increasing $V_{BG}$. At low $V_{BG}$ (region I: $V_{BG} < -12.5$ V), we observe a linear ohmic-like $I_{DS}$-$V_{DS}$ relation. At the intermediate $V_{BG}$ range (region II: $-12.5 \text{ V} \leq V_{BG} \leq 0 \text{ V}$), $I_{DS}$ exhibits the rectified, diode-like $I_{DS}$-$V_{DS}$ behavior. Further increasing $V_{BG}$ to region III recovers the linear conduction.

To understand the origin of the evolved $I_{DS}$-$V_{DS}$ characteristic, we superimposed on Fig. 3(c) the transfer curves of this sample in the mono-domain states. The corresponding transition voltages are $V_t = -13$ V for $P_{down}$ and $V_t = 0$ V for $P_{up}$ state, which are in excellent agreement with the boundary $V_{BG}$ voltages that separate regions I, II and III, suggesting that it is the band alignment between the $P_{up}$ and $P_{down}$ sides of the channel that determines the conduction characteristic. When the sample is gated into region I, $E_F$ for both sides is deep in the gap region, leading to high channel resistance. We expect an intrinsic to $n$-doped type interface, and the channel conduction can be modeled as two resistors $R_{down}$ and $R_{up}$, corresponding to the $P_{down}$ and $P_{up}$ domains, respectively, in series connection. In region II, $E_F$ for the $P_{down}$ side of the channel is approaching the conduction band edge, so MoS$_2$ is doped with a high density of free electrons in the extended states at room temperature, while the $P_{up}$ side of the channel remains in the semiconducting state. We thus expect a Schottky junction forming in the vicinity of the DW, with the effective barrier height $\Phi_B^{eff}$ corresponding to the difference in $E_F$ between the $P_{up}$ and $P_{down}$ sides (Fig. 4(a) inset). Similarly, well within region III, both sides of the channel are doped with high densities of electrons, yielding low channel resistance with linear conduction.

In region II, we need to consider the additional junction contribution in series connection with $R_{up}$ and $R_{down}$ to model the total channel conduction. We quantitatively modeled the rectified $I$-$V$ in region II using the thermionic emission model [13, 47]:

$$I_{DS} = A_{2D}^* T^{3/2} W \exp\left(-\frac{e\Phi_B^{eff}}{k_B T}\right) \exp\left(\frac{eV_{DS}}{nk_B T}\right) \left[1 - \exp\left(-\frac{eV_{DS}}{k_B T}\right)\right] \quad (3)$$

with $A_{2D}^* = \left[(eM_c k_B^{3/2})/(\pi \hbar^2)\right](m_t^*/2\pi)^{1/2}$,

where $W$ is the width of the conduction channel, $n$ is the ideality factor, which is normally between 1 and 2, and $m_t^* = 0.45 m_0$ is the transverse electron effective mass for ML MoS$_2$ [3]. Here $A_{2D}^*$ is the 2D effective Richardson constant, with the number of equivalent conduction band minima $M_c = 2$. Figures 4(a)-(b) show the $I_{DS}$-$V_{DS}$ curves taken on sample D5 in region II (between $V_{BG} = -12$ V and 0 V for this sample). At $V_{BG} = -10$ V, the channel conduction can be well described by Eq. (3) with $\Phi_B^{eff} = 0.51$ eV, which can be further tuned by $V_{BG}$. The domain patterning induced junction states have been observed in all four samples investigated (D3-D6), with the ideality factor $n$ ranging from 1.4 to 1.8, consistent with previously reported results [9, 12].

The fact that Eq. (3) can give an accurate description of the conduction characteristic indicates that the resistive contribution of $R_{down}$ and $R_{up}$ is also absent at $V_{BG} = -10$ V. As MoS$_2$ exhibits more than one order of magnitude change in conductance between the two polarization states in region II (Figs. 1(b) and 3), it is reasonable to neglect $R_{down}$. The absence of $R_{up}$, on the other hand, suggests that the extension of the junction transition region is comparable with or larger than the $P_{up}$ side channel length at low carrier density. At sufficiently high $V_{BG}$ ($V_{BG} \geq -6$ V), when the junction width decreases with increasing carrier density and becomes shorter than the associated channel length, Eq. (3) can no longer provide a satisfactory fit to $I_{DS}$-$V_{DS}$. Rather, the conduction has to be modeled as a Schottky junction in series connection with a resistor [31]:

$$V'_{DS} = V_{DS}(I_{DS}) + I_{DS} R_{up}, \qquad (4)$$

where $V_{DS}(I_{DS})$ is defined by Eq. (3). Figure 4(b) shows the fit for the $I_{DS}$-$V_{DS}$ curve at $V_{BG} = -4$ V to this model, where the first term yields a barrier height of $\Phi_B^{eff} = 0.43$ eV. Comparing the extracted $R_{up}$ with the channel resistance in the uniform $P_{up}$ state yields a channel length of 23 nm, much shorter than the associated $P_{up}$ domain length (350 nm), confirming that on the $P_{up}$ side the junction transition region extends well beyond the vicinity of the DW.

As a control experiment, we erased the domain structure by thermally heating the sample in vacuum to 80 °C, close to its Curie temperature, for 2 hours. The copolymer film then recovers its as-deposited state with polarization randomly orientated, which can be attributed to its low DW energy [38]. After thermal depolarization, the $I_{DS}$-$V_{DS}$ curve at $V_{BG} = -10$ V agrees well with the linear behavior in the as-deposited state (Fig. 4(c)), indicating that the observed junction state is resulting from the ferroelectric domain patterning and the sample quality is preserved during the polarization switching.

As the Schottky barrier height $\Phi_B^{eff}$ is defined by the band alignment between the $P_{down}$ and $P_{up}$ states (Fig. 4(a) inset), it can be further tuned by $V_{BG}$. Figure 4(d) shows $\Phi_B^{eff}$ as a function of the normalized back-gate voltage ($V_{BG}$-$V_{mid}$)/$V_W$ for samples D3-D6, where $V_{mid}$ and $V_W$ are the middle point and width of region II, respectively. All samples exhibit consistent $V_{BG}$-dependence of $\Phi_B^{eff}$, which decreases with increasing $V_{BG}$ by 90-150 meV. This reduction in $\Phi_B^{eff}$ is a consequence of the high density of states close to the MoS$_2$ conduction band edge, which results in a moderate change of $E_F$ on the $P_{down}$ side where the sample is doped close to the conduction band.

At the low $V_{BG}$ end of region III, close to the boundary with region II, the channel conduction still exhibits slight nonlinearity but can no longer be properly modeled by Eq. (4) [31], even though the $P_{up}$ state $E_F$ is still about 0.4-0.5 eV below the extended states (Fig. 4(d)). This can be well explained by the band tail states described by Eq. (1), which would limit the gate-tunable range of $\Phi_B^{eff}$, or the width of region II ($V_W$). As shown in Ref. [6], these impurity states can extend to up to 0.4-0.5 eV below the conduction band of MoS$_2$, and can contribute to hopping-type conduction, which naturally explains the lack of a well-defined energy barrier at this gating range. Overall, a range of $\Phi_B^{eff}$ from 0.38 eV to 0.57 eV has been realized in our samples (Fig. 4(d)).

In conclusion, we have demonstrated that nanoscale ferroelectric domain patterning can be utilized to reversibly tune the functionality of a 2D channel, while polarization switching does not have a pronounced impact on the channel mobility. This approach can be applied to a wide range of van der Waals materials to design various homojunctions and nanostructures, such as nanoribbons, nanodots and superlattices. Compared to the lithographical methods, the nonvolatile field effect allows one to impose different functional designs on the same 2D channel without introducing disordered interface/edge states. It also opens up the opportunity to integrate logic, memory, and photovoltaic functionalities in a single material platform for nanoelectronic and optoelectronic applications.

**Acknowledgments**

We would like to thank Zhixian Zhou for insightful discussions, Dawei Li and Peter Kosch for experimental assistance, and Yongfeng Lu for the access to the Raman system. This work was supported by the National Science Foundation (NSF) through Nebraska Materials Research Science and Engineering Center (MRSEC) (Grant No. DMR-1420645) (ferroelectric polymer deposition and device fabrication), NSF CAREER Grant No. DMR-1148783 (electrical and optical characterizations) and NSF Grant No. OIA-1538893 (scanning probe studies), and the U.S. Department of Energy (DOE), Office of Science, Basic Energy Sciences (BES), under Award No. DE-SC0016153 (low temperature studies and data modeling). The research was performed in part in the Nebraska Nanoscale Facility: National Nanotechnology Coordinated Infrastructure and the Nebraska Center for Materials and Nanoscience, which are supported by NSF under Award ECCS: 1542182, and the Nebraska Research Initiative.

# Figure 1

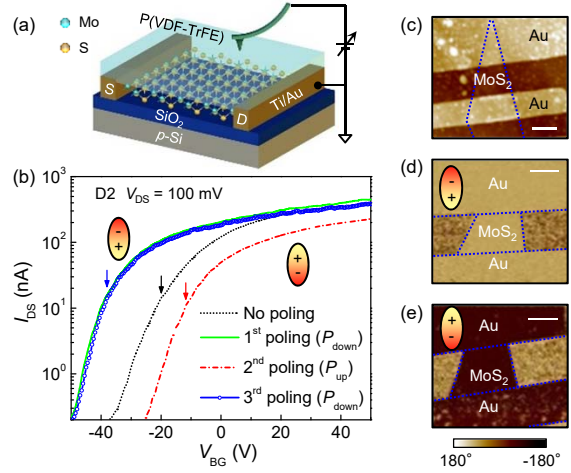

FIG. 1. (a) Device schematic. (b) Room-temperature transfer characteristics for sample D2 in the initial no poling state, and after writing P(VDF-TrFE) with +10 V (1st poling), -10 V (2nd poling), and +10 V (3rd poling) tip bias. The arrows mark the corresponding $V_t$s. (c) The AFM topography image of the sample and the PFM phase images after (d) the 1st and (e) 2nd poling. The scale bars are 1 μm. The dotted lines outline the MoS$_2$ flake and the Au electrodes.

Figure 2

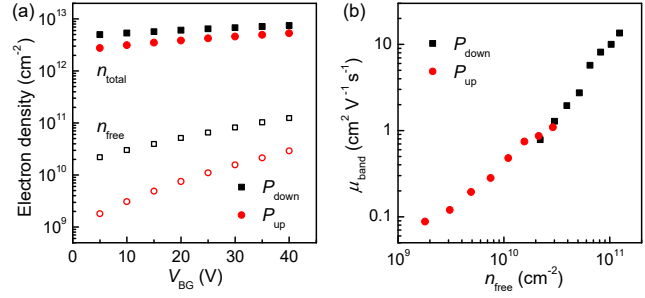

FIG. 2. (a) Calculated $n_{total}$ (solid symbols) and $n_{free}$ (open symbols) as a function of $V_{BG}$ for sample D3 at $V_{BG} > V_t$ for both polarization states, and (b) the corresponding $\mu_{band}$ as a function of $n_{free}$.

Figure 3

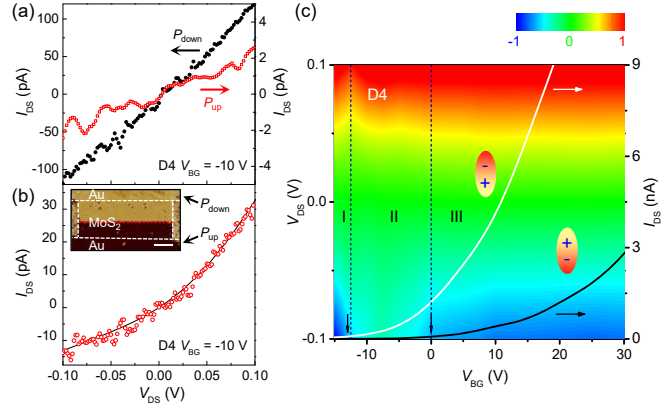

FIG. 3. (a) Room-temperature $I_{DS}$-$V_{DS}$ of sample D4 at $V_{bg}$ = -10 V for the mono-domain states, and (b) the half $P_{up}$-half $P_{down}$ state (open symbols) with a fit to Eq. (3) (solid line). Inset: PFM phase image of the domain structure. Scale bar: 1 μm. (c) Normalized $I_{DS}$ vs. $V_{DS}$ and $V_{BG}$ for the half $P_{up}$-half $P_{down}$ state. Superimposed on the plot are the transfer curves at $V_{DS}$ = 100 mV for the uniform $P_{up}$ and $P_{down}$ states (right axis). The vertical arrows mark the $V_t$s. The dashed lines mark the boundaries separating regions I, II, and III.

# Figure 4

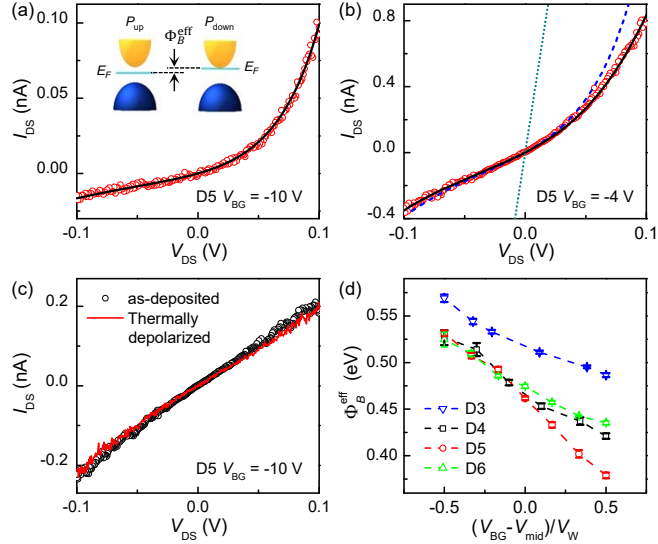

FIG. 4. (a) $I_{DS}$-$V_{DS}$ (open symbols) at $V_{BG}$ = -10 V with a fit to Eq. (3) (solid line). Inset: Schematic band diagram for MoS$_2$ in the $P_{up}$ and $P_{down}$ states. (b) $I_{DS}$-$V_{DS}$ (open symbols) at $V_{BG}$ = -4 V with a fit to Eq. (4) (solid line). The dashed and dotted lines correspond to the first and second terms in Eq. (4), respectively. (c) $I_{DS}$-$V_{DS}$ at $V_{BG}$ = -10 V taken in the as-prepared state of P(VDF-TrFE) (black symbols) and after domain patterning and thermal depolarization (red line). (d) $\Phi_B^{eff}$ as a function of normalized $V_{BG}$ for samples D3-D6.